\documentclass[12pt]{article}

\usepackage{hyperref}
\usepackage{amsmath}
\usepackage{graphicx}
\usepackage{cite}

\makeatletter

\@addtoreset{equation}{section}

\def\appendix{\par
 \setcounter{section}{0}
 \setcounter{subsection}{0}
 \setcounter{equation}{0}
 \def\thesection{\Alph{section}}}

\makeatother

\newcommand{\ga}{\gamma}
\newcommand{\de}{\delta}

\newcommand\veps{\varepsilon}

\newcommand{\si}{\sigma}

\newcommand{\Ga}{\Gamma}

\newcommand{\Si}{\Sigma}

\newcommand{\ri}{{\mathrm{i}}}
\newcommand{\re}{{\mathrm{e}}}

\newcommand{\rT}{{\mathrm{T}}}
\newcommand{\rR}{{\mathrm{R}}}
\newcommand{\rS}{{\mathrm{S}}}
\newcommand{\rL}{{\mathrm{L}}}
\newcommand{\rV}{{\mathrm{V}}}
\newcommand{\rd}{{\mathrm{d}}}
\newcommand{\rw}{{\mathrm{w}}}

\renewcommand{\L}{{\cal L}}

\newcommand{\ren}{{\mathrm{R}}}
\newcommand{\sren}{{\mathrm{SR}}}

\def\mathswitch#1{\relax\ifmmode#1\else$#1$\fi}
\def\mathswitchr#1{\relax\ifmmode{\mathrm{#1}}\else$\mathrm{#1}$\fi}
\def\mathswitchit#1{\relax\ifmmode{#1}\else$#1$\fi}

\newcommand{\PW}{\mathswitchr W}
\newcommand{\PZ}{\mathswitchr Z}

\newcommand{\Pf}{f}

\newcommand{\MZ}{\mathswitch {M_\PZ}}

\newcommand{\scrs}{\scriptscriptstyle}

\newcommand{\sw}{\mathswitch {s_{\scrs\PW}}}
\newcommand{\cw}{\mathswitch {c_{\scrs\PW}}}

\newcommand{\Qf}{\mathswitch {Q_\Pf}}

\def\refeq#1{\mbox{(\ref{#1})}}

\def\reffi#1{\mbox{Fig.~\ref{#1}}}

\def\refse#1{\mbox{Section~\ref{#1}}}

\def\citere#1{\mbox{Ref.~\cite{#1}}}
\def\citeres#1{\mbox{Refs.~\cite{#1}}}
\def\nn{\nonumber}

\newcommand{\GaBFM}{\hat\Gamma}
\newcommand{\FA}{A}
\newcommand{\FZ}{Z}
\newcommand{\FV}{V}
\newcommand{\FAhat}{\hat A}
\newcommand{\FZhat}{\hat Z}
\newcommand{\FVhat}{\hat V}
\newcommand{\Ff}{f}
\newcommand{\Ffbar}{\bar{f}}

\newcommand*\beq{\begin{equation}}
\newcommand*\eeq{\end{equation}}
\newcommand*\beqar{\begin{eqnarray}}
\newcommand*\eeqar{\end{eqnarray}}
\newcommand*\barr{\begin{array}}
\newcommand*\earr{\end{array}}
\newcommand*\bfi{\begin{figure}}
\newcommand*\efi{\end{figure}}
\newcommand*\btab{\begin{table}}
\newcommand*\etab{\end{table}}
\newcommand*\bpm{\begin{pmatrix}}
\newcommand*\epm{\end{pmatrix}}

\makeatletter
\newcommand\myparagraph{\@startsection{paragraph}{4}{\z@}%
  {-10\p@ \@plus 6\p@ \@minus 3\p@}%
  {3\p@}%
  {\normalfont\itshape}%
}
\def\dsl{\mathpalette\make@slash}
\def\make@slash#1#2{\setbox\z@\hbox{$#1#2$}%
  \hbox to 0pt{\hss$#1/$\hss\kern-\wd0}\box0}
\makeatother

\begin{document}

\thispagestyle{empty}
\def\thefootnote{\fnsymbol{footnote}}
\setcounter{footnote}{1}
\null
\strut\hfill FR-PHENO-2021-001
\vskip 0cm
\vfill
\begin{center}
{\large \bf Electric charge renormalization to all orders
\par} \vskip 2.5em
{\large
{\sc Stefan Dittmaier
}\\[1ex]
{\normalsize
\it
Albert-Ludwigs-Universit\"at Freiburg,
Physikalisches Institut, \\
Hermann-Herder-Stra\ss{}e 3,
D-79104 Freiburg, Germany 
}
}

\par \vskip 1em
\end{center} \par
\vskip 3cm
{\bf Abstract:} \par
The electric charge renormalization constant, as defined in the Thomson limit,
is expressed in terms of self-energies
of the photon--Z-boson system in an arbitrary $R_\xi$-gauge to all perturbative
orders. The derivation as carried out in the Standard Model
holds in all spontaneously broken gauge theories with the
SU(2)$_\rw\times$U(1)$_Y$ gauge group in the electroweak sector
and is based on the application of charge universality 
to a fake fermion with infinitesimal weak hypercharge and vanishing weak isospin,
which effectively decouples from all other particles.
Charge universality, for instance, follows from the 
known universal form of the charge renormalization constant as derived within the 
background-field formalism.
Finally, we have generalized the described procedure to gauge theories
with gauge group U(1)$_Y{\times}G$ with any Lie group $G$, only assuming
that electromagnetic gauge symmetry is unbroken and mixes with
U(1)$_Y$ transformations in a non-trivial way.
\par
\vfill
\noindent
January 2021 \par
\vskip .5cm
\null
\setcounter{page}{0}
\clearpage
\def\thefootnote{\arabic{footnote}}
\setcounter{footnote}{0}

\section{Introduction}

The issue of electric charge renormalization is as old as quantum electrodynamics (QED)
and relativistic quantum field theory.
In order to give the electric unit charge~$e$ the same physical meaning 
as in classical electrodynamics, in QED the renormalized value of~$e$ is defined
by the condition that the electron--photon vertex for physical (on-shell)
electrons does not receive any corrections in the limit of zero-momentum transfer
for the photon, also known as {\it Thomson limit}. 
Gauge invariance and this condition, in particular, imply that
the low-energy limit of Compton scattering goes over into classical
Thomson scattering without receiving radiative corrections---a 
statement also known as {\it Thirring's theorem}~\cite{Thirring:1950cy}.
By virtue of the famous Ward identity for the electron--photon vertex, the 
Thomson condition implies that the product $eA^\mu$ of $e$ and the 
(canonically normalized) photon field $A^\mu$
is not renormalized, i.e.\ $e_0A_0^\mu = eA^\mu$ if we denote {\it bare}
quantities before renormalization with a subscript zero.
Following the usual approach of multiplicative renormalization,
the bare and renormalized quantities are related by
$e_0=Z_e e$ and $A_0^\mu = Z_{AA}^{1/2}A^\mu$, so that the charge
renormalization constant $Z_e$ and the photon wave-function renormalization
constant $Z_{AA}$ are related by $Z_e=Z_{AA}^{-1/2}$.
In practice, this means that $Z_e$ can be determined upon calculating
the photon self-energy only, although it is defined via some condition
demanded for an interaction vertex. 
All this is standard knowledge and very well described in many
textbooks on QFT, such as 
\citeres{Peskin:1995ev,Weinberg:1996kr,Bohm:2001yx,Schwartz:2013pla}.

In the Standard Model (SM) of particle physics and its extensions,
the unit charge~$e$ is defined by the same condition in the Thomson limit
as in QED, but owing to the more complicated gauge symmetry as compared to 
QED and the mixing between photons and Z~bosons, 
the determination of $Z_e$ from this condition is way more complicated.
In the 1980s and early 1990s the renormalization of the (electroweak part 
of the) SM was formulated in different 
variants~\cite{Ross:1973fp,Sirlin:1980nh,Bardin:1980fe,Fleischer:1980ub,%
Aoki:1982ed,Bohm:1986rj,Hollik:1988ii,Denner:1991kt}
and worked out in detail at the one-loop level
(see \citere{Denner:2019vbn} for a detailed review and further references).
Using again a Ward identity for the fermion--photon vertex, it was possible
to express the one-loop contribution to $Z_e$ in terms of self-energies
of the photon--Z-boson system, but the underlying Ward identity was first 
justified by explicit one-loop calculations and derived from the
underlying Lee identities for vertex functions in \citere{Denner:2019vbn}
only recently.
Whether it is possible to derive $Z_e$ in all perturbative orders
from self-energies only, was (to our knowledge) not clear until the
mid 1990s.

The first all-order prescription to express $Z_e$ in terms of
self-energies was formulated in \citere{Denner:1994xt} within
the background-field method 
(BFM)~\cite{DeWitt:1967ub,tHooft:1975uxh,DeWitt:1980jv,Boulware:1980av,%
Abbott:1980hw,Abbott:1981ke} 
(see also \citeres{Weinberg:1996kr,Bohm:2001yx,Schwartz:2013pla}),
which provides an alternative to the conventional 
formalism for quantizing gauge theories.
The special feature of the BFM is the gauge invariance of the 
effective action which implies QED-like Ward identities for vertex functions.
Applied to the fermion--photon vertex, these identities can be used
to show that in analogy to QED the combination $eA^\mu$ is not renormalized,
so that $Z_e$ is determined by the photon wave-function renormalization
constant of the BFM. Even Thirring's theorem could be proven to all
orders with arguments based on BFM gauge 
invariance~\cite{Dittmaier:1997dx}.
Exploiting the simple connection between the charge renormalization constant
$Z_e$ and the photon self-energy, the constant $Z_e$ was explicitly calculated
at the two-loop level in \citere{Degrassi:2003rw}. In the same paper it was also
verified that the explicit two-loop result for $Z_e$, as obtained in the BFM, leads 
to the correct Thomson limit of the renormalized fermion--photon vertex
in the conventional 't~Hooft--Feynman gauge, thereby confirming the 
universality (in the sense of independence of gauge-fixing procedures or
quantization formalisms) of $Z_e$, which follows from the fact that the
Thomson condition is formulated at the basis of a physical S-matrix element.

Existing electroweak calculations beyond one loop are still scarce,
including for instance the full ${\cal O}(\alpha^2)$ corrections to muon 
decay~\cite{Freitas:2002ja,Awramik:2002vu}
and the fermion-loop contributions of ${\cal O}(\alpha^2)$~\cite{Freitas:2013dpa}
and ${\cal O}(\alpha_{\mathrm{s}}\alpha^2)$~\cite{Chen:2020xot}
to the Z-boson decay.
Since those calculations have not been carried out in the BFM, 
the charge renormalization constant was calculated
within the conventional quantization formalism. 
For the calculation of the required higher-order contributions to $Z_e$
the authors of \citeres{Freitas:2002ja,Awramik:2002vu,Freitas:2013dpa,Chen:2020xot}
refer to \citere{Bauberger:1997zz}, where an all-order result for $Z_e$
in terms of gauge-boson self-energy contributions is given for 
$R_\xi$-gauges. Inspecting, however, the derivation of $Z_e$ of
\citere{Bauberger:1997zz}, which is based on the Slavnov--Taylor (ST)
identities for the Green functions of the photon--fermion vertex and
the propagators in the photon--Z-boson sector, we find severe inconsistencies,
as detailed in the Appendix.
At the two-loop level, however, the claimed form of $Z_e$ was confirmed by an explicit
calculation in 't~Hooft--Feynman gauge in
\citeres{Actis:2006ra,Actis:2006rb,Actis:2006rc}, where it was shown that the sum
of all genuine vertex corrections and fermionic wave-function corrections to
the photon--fermion vertex vanishes in the Thomson limit;
this is exactly the part in the calculation of $Z_e$ that is ruled by 
gauge invariance and that is yet unproven in conventional $R_\xi$-gauge
to all orders.

The purpose of this paper is to fill this gap and to derive an all-order
form for $Z_e$ in arbitrary $R_\xi$-gauge. 
To anticipate our result, we confirm the previously claimed all-order form of $Z_e$,
thus, providing an a posteriori justification to the charge renormalization
as carried out in the calculations of 
\citeres{Freitas:2002ja,Awramik:2002vu,Freitas:2013dpa,Chen:2020xot}.
Our proof is based on {\it charge universality}, which is the statement that
any fermion may be taken in the Thomson renormalization condition for the
fermion--photon vertex without changing the result on
the renormalized unit charge~$e$.
Charge universality, in particular, implies that the ratios between bare
charges of particles, which are often fixed by symmetry relations
(like the charge ratio of up and down quarks), survive the procedure
of renormalization.
In \citere{Aoki:1982ed} charge universality was proven to all orders by
employing Lee identities to appropriate vertex functions;%
\footnote{Actually, \citere{Aoki:1982ed} offers three derivations of
charge universality. 
The first version (Sect.~3.4.2 of \citere{Aoki:1982ed}) is based on
arguments of S-matrix theory, but leaves some loop holes as pointed out also
there; it is thus more a proof of self-consistency.
The second and third versions of the proof, which are based on Lee identities (Sect.~3.4.3)
and Becchi--Rouet--Stora invariance (Sect.~3.4.4) of the theory, respectively, 
make use of the Landau gauge in some steps. This gauge choice does not
restrict the proof of the property of charge universality, but the
derived relations between Green functions, renormalization constants, etc.\ 
do not all hold in an arbitrary $R_\xi$-gauge.}
alternatively charge universality follows from the derivation~\cite{Denner:1994xt}
of $Z_e$ in the BFM, which does not distinguish any fermion.
With charge universality proven, we can make use of this property at will.
We are, thus, allowed to demand the Thomson renormalization condition 
to fix $e$ even for a ``fake fermion'' that does not exist in the SM as long as it
does not change any prediction for observables.
This decoupling property is guaranteed by taking a non-chiral fake fermion with
an infinitesimal weak hypercharge and vanishing weak isospin. 
As will be shown, all non-trivial irreducible vertex corrections and
fermionic wave-function contributions appearing in the 
Thomson renormalization condition for the fake fermion drop out in the
calculation of $Z_e$, and only the contributions from the self-energies
of the photon--Z-boson system remain.
Finally, we mention that decoupling ``fake particles'' were already used
in the formulation of on-shell renormalization conditions for mixing angles
in \citere{Denner:2018opp}. 
This concept appears very promising whenever it is desirable to reduce
unnecessary dependences on specific particles or model parameters from 
renormalization procedures.

Since the described procedure only makes use of the gauge structure of the SM,
the SM result for $Z_e$ in terms of renormalization constants for the 
photon--Z-boson system
holds in all gauge theories with the
SU(2)$_\rw\times$U(1)$_Y$ gauge group and the same pattern of 
spontaneous gauge symmetry breaking as the SM in the electroweak sector.
The further generalization of the charge renormalization procedure
to spontaneously broken gauge theories
with gauge group U(1)$_Y{\times}G$ with any Lie group $G$ and an 
embedding of unbroken electromagnetic gauge symmetry analogous to the SM
is fully straightforward and presented after our treatment of the SM.

The article is organized as follows:
In \refse{se:OSren} we review the on-shell renormalization of the photon--Z-boson
system as far as relevant for treating external photons in scattering
amplitudes, in order to introduce the renormalization constants that are
relevant for charge renormalization and to keep this paper self-contained
as much as possible. In this context we also consider the implications
of electromagnetic gauge invariance on those renormalization constants
and discuss the masslessness of the photon, both in conventional
$R_\xi$-gauge and the BFM.
Section~\ref{se:chargerenrxi} describes charge renormalization and 
the determination of $Z_e$ in the BFM in detail,
following the proposal sketched in \citere{Denner:1994xt};
this derivation can also be seen
as an independent proof of charge universality.
Section~\ref{se:chargerenrxi} presents the all-order derivation of
$Z_e$ in conventional $R_\xi$-gauge based on the idea of introducing
a fake fermion and constitutes the major part of this work.
The generalization of the charge renormalization procedure to
U(1)$_Y{\times}G$ gauge theories is described in \refse{se:BSM}.
Finally, we give some conclusions in \refse{se:concl}.

\section{On-shell renormalization in the photon--Z-boson sector}
\label{se:OSren}

\subsection{Basic definitions and on-shell renormalization conditions}
\label{se:VVdefs}

We first consider on-shell (OS) renormalization in the photon--Z-boson sector
of the SM
to all orders and work it out to the extent to which 
it is important in the formulation of charge renormalization.
The procedure is widely identical in the conventional
formalism and in the BFM, so that we keep
the presentation generic in the first steps.
Unless stated otherwise, we consistently make use of the conventions and notation 
of \citere{Denner:2019vbn} for all field-theoretical quantities.

We start from 
the decomposition of the {\it unrenormalized} two-point functions
$\Gamma^{\FV^{\prime}\FV}_{\mu\nu}$ in momentum space into transversal (T)
and longitudinal (L) parts,
\begin{align}\label{eq:Lorentz_VV}
\Gamma^{\FV^{\prime}\FV}_{\mu \nu }(-k,k) ={}&
\left(g_{\mu \nu} -\frac{k_{\mu} k_{\nu }}{k^{2}}\right)
\Gamma^{\FV^{\prime}\FV}_{\rT}(k^{2})
+\frac{k_{\mu} k_{\nu }}{k^{2}}
\Gamma^{\FV^{\prime}\FV}_{\rL}(k^{2}),  \qquad V,V' = A,Z,
\end{align}
where $k$ is the momentum transfer.
{\it Unrenormalized} means that no 
renormalization transformation of parameters and fields
is performed yet and the vertex functions 
are obtained from taking functional derivatives of the effective action
with respect to bare fields and expressed
in terms of bare parameters. Bare parameters and fields are
marked by a subscript ``0'' in the following.

In the renormalization procedure, mainly
the transversal parts $\Gamma^{\FV^{\prime}\FV}_{\rT}$ will be
relevant.
The unrenormalized vertex function $\Gamma^{\FV^\prime\FV}_{\rT}$
receives lowest-order contributions
and higher-order corrections from one-particle-irreducible (1PI)
loop diagrams and tadpole contributions.%
\footnote{Following the conventions of \citere{Denner:2019vbn} for one-loop
corrections, we define self-energy functions like
$\Si^{\FV^\prime\FV}$ to contain all contributions from
tadpole diagrams and tadpole counterterms, here generically called
$\Si^{\FV^\prime\FV}_{\mathrm{tad}}$. 
The notation is somewhat at variance from \citere{Denner:2019vbn},
where (one-loop) tadpole counterterm contributions are separated from the
explicit tadpole loops and called $\Si^{\FV^\prime\FV}_{\de t}$.
Collecting all contributions with explicit tadpole diagrams and/or tadpole
counterterms into $\Si^{\FV^\prime\FV}_{\mathrm{tad}}$
saves us from some clutter and an arbitrary classification of diagrams
with both tadpole loops and tadpole counterterms.
The tadpole parts $\Si^{\FV^\prime\FV}_{\mathrm{tad}}$
depend on the tadpole scheme, i.e.\ on the details of the definition
of vacuum expectation values of Higgs fields 
(see, e.g., \citere{Denner:2019vbn} and references therein), but those details
will not play a role in the following.}
Considering the definition of potentially non-zero vacuum expectation values 
as part of defining the theory in terms of bare quantities, the
tadpole counterterms are also part of $\Gamma^{\FV^\prime\FV}_{\rT}$.
Splitting off the lowest-order parts from $\Gamma^{\FV^\prime\FV}_{\rT}$
and calling the higher-order part the unrenormalized 
self-energy $\Si^{\FV^\prime\FV}_{\rT}$, we have
\begin{align}
\label{eq:GaVVunren}
\Gamma^{\FV^\prime\FV}_{\rT} (k^2) ={}
-\de_{\FV^\prime\FV} \, (k^2 -\de_{\FV\FZ} \,M_{\PZ,0}^2)
-\Si^{\FV^\prime\FV}_{\rT}(k^2,\{c_{i,0}\}),
\end{align}
with $\de_{\FV^\prime\FV}$ and $\de_{\FV\FZ}$ being a Kronecker deltas and 
$M_{\PZ,0}$ denoting the bare Z-boson mass and
\begin{align}
\label{eq:SiVV0}
\Si^{\FV^\prime\FV}_{\rT}(k^2,\{c_{i,0}\}) ={}
\Si^{\FV^\prime\FV}_{\rT,\mathrm{1PI}}(k^2,\{c_{i,0}\}) +
\Si^{\FV^\prime\FV}_{\rT,\mathrm{tad}}(k^2,\{c_{i,0}\}).
\end{align}
Here, $\Si^{\FV^\prime\FV}_{\rT,\mathrm{1PI}}$ comprises the
1PI loop diagrams and
$\Si^{\FV^\prime\FV}_{\rT,\mathrm{tad}}$ all contributions 
containing tadpole corrections.
The list $\{c_{i,0}\}$ of arguments is added in order to make clear
that the self-energies are parametrized by the bare parameters $c_{i,0}$
of the theory.

In the OS renormalization scheme,
the renormalization transformation for the photon and Z-boson fields is
given by
\begin{align}\label{eq:VVfieldren}
\bpm \FZ_{0} \\ \FA_{0} \epm ={}
\bpm Z_{\FZ\FZ}^{1/2} & Z_{\FZ\FA}^{1/2}  \\[1ex]
     Z_{\FA\FZ}^{1/2} & Z_{\FA\FA}^{1/2} \epm
\bpm \FZ \\ \FA \epm,
\end{align}
where the bare fields on the l.h.s.\ are expressed in terms of the renormalized fields
$Z$, $A$ on the r.h.s.\ and $Z_{\FV'\FV}$
are the field renormalization constants to be determined by the OS
renormalization conditions.
Those conditions are demanded for the {\it renormalized} two-point functions
$\Ga^{\FV'\FV}_{\rR,\mu\nu}$, which are related to the
unrenormalized two-point functions according to
\begin{align}
\label{eq:GaVVrentrafo}
\Ga^{\FV^{\prime}\FV}_{\rR,\mu \nu }(-k,k) ={}
\sum_{V_1,V_2=A,Z} Z_{\FV_1\FV^{\prime}}^{1/2} Z_{\FV_2\FV}^{1/2}
\Ga^{\FV_1\FV_2}_{\mu \nu }(-k,k),
\end{align}
which directly follows from the field transformation \refeq{eq:VVfieldren}.
The renormalization transformation of the remaining fields and of the
parameters of the theory does not spoil this relation, because the other
field transformations merely redistribute terms between vertex and 
internal propagator corrections, and the parameter renormalization
transformation merely reparametrizes the vertex functions.
Obviously the form of relation \refeq{eq:GaVVrentrafo} 
carries over to the transversal and
longitudinal parts of $\Ga^{\FV^{\prime}\FV}_{\rR,\mu\nu}$ and
$\Ga^{\FV^\prime\FV}_{\mu\nu}$ independently.
Moreover, there is the obvious symmetry
\begin{align}
\Ga^{\FV^{\prime}\FV}_{\rR,\mu \nu }(-k,k) ={}
\Ga^{\FV\FV^{\prime}}_{\rR,\mu \nu }(-k,k), \qquad
\Ga^{\FV^{\prime}\FV}_{\mu \nu }(-k,k) ={}
\Ga^{\FV\FV^{\prime}}_{\mu \nu }(-k,k). 
\end{align}

For charge renormalization we need the constants $Z_{\FA\FA}$
and $Z_{\FZ\FA}$, which are derived from the renormalization
conditions for on-shell ($k^2=0$) photons,
\begin{align}
\label{eq:GaVArencond}
\lim_{k^2\to0} \Gamma^{\FZ\FA}_{\ren,\mu\nu} (-k,k) \, \veps^{\nu }(k) ={}& 0,
\\
\label{eq:dGaVArencond}
\lim_{k^2\to0} \frac{1}{k^2}\, \Gamma^{\FA\FA}_{\ren,\mu\nu}(-k,k) \,\veps^{\nu}(k)
={}& -\veps_{\mu}(k),
\end{align}
where $\veps^{\mu}(k)$ is the polarization vector of a photon with momentum~$k$.
The first of those conditions ensures that on-shell photons do
not fluctuate into Z-boson states, the second keeps photon states canonically
normalized, i.e.\ normalized as in lowest order.
Note that there is no extra condition to fix the pole in the photon
propagator to the location at $k^2=0$, because there is no free parameter
like a photon mass that could be renormalized to achieve this.
This condition is implied by gauge invariance automatically.
However, this fact is encoded in the conventional formalism
and in the BFM in different ways and will be discussed in the subsequent sections.

The conditions $\refeq{eq:GaVArencond}$ and \refeq{eq:dGaVArencond}
involve only the transversal parts of $\Gamma^{\FV\FA}_{\ren,\mu\nu}$,
while the longitudinal parts 
drop out in these relations.
To derive the implications on the transversal parts and on the desired
renormalization constants, we decompose the renormalized transversal parts of the
vertex functions into loop contributions and 
remainders that contain lowest-order and counterterm
contributions,
\begin{align}
\label{eq:GaVVren}
\Gamma^{\FV^\prime\FV}_{\rR,\rT} (k^2) 
={}&
-\de_{\FV^\prime\FV} \, (k^2 -\de_{\FV\FZ} \,\MZ^2)
-\Si^{\FV^\prime\FV}_{\ren,\rT}(k^2,\{c_i\})
\nn\\
={}&
- Z^{1/2}_{\FA\FV^\prime} Z^{1/2}_{\FA\FV} k^2
- Z^{1/2}_{\FZ\FV^\prime} Z^{1/2}_{\FZ\FV}(k^2-\MZ^2-\de\MZ^2)
-\Si^{\FV^\prime\FV}_{\sren,\rT}(k^2,\{c_i\}),
\end{align}
where $\{c_i\}$ indicates the parametrization in terms of renormalized 
parameters $c_i$.
The renormalized self-energies $\Si^{\FV^\prime\FV}_{\ren,\rT}$ 
comprise all higher-order corrections (loops, tadpoles, counterterms, and
mixed contributions thereof) and are UV~finite by construction.
The terms in the second line containing the $Z$ factors directly result from the
free Lagrangian after the field transformation \refeq{eq:VVfieldren} and from the 
Z-boson mass renormalization $M_{\PZ,0}^2=\MZ^2+\de\MZ^2$, which will not be 
important in the following. 
The {\it subgraph-renormalized} (SR) self-energies $\Si^{\FV^\prime\FV}_{\sren,\rT}$ 
contain all loop and tadpole contributions and insertions of counterterms into
loops, but no genuine counterterm contributions without loop part which are extracted
by the terms with the $Z$~factors. The SR self-energies are, in general, not
UV~finite, but the potential UV~divergences are of polynomial structure
in $k^2$ with degree one.
Using \refeq{eq:GaVVren}, we can express the renormalized self-energies
in terms of SR self-energies and renormalization constants,
\begin{align}
\Si^{\FV^\prime\FV}_{\ren,\rT}(k^2,\{c_i\})
={}&
\Si^{\FV^\prime\FV}_{\sren,\rT}(k^2,\{c_i\})
+ \left( Z^{1/2}_{\FA\FV^\prime} Z^{1/2}_{\FA\FV}
+ Z^{1/2}_{\FZ\FV^\prime} Z^{1/2}_{\FZ\FV}
-\de_{\FV^\prime\FV} \right) k^2
\nn\\ & {}
- \left( Z^{1/2}_{\FZ\FV^\prime} Z^{1/2}_{\FZ\FV}
-\de_{\FV^\prime\FZ} \, \de_{\FV\FZ} \right) \MZ^2
- Z^{1/2}_{\FZ\FV^\prime} Z^{1/2}_{\FZ\FV} \, \de\MZ^2.
\end{align}
Recalling further \refeq{eq:VVfieldren} and \refeq{eq:GaVVrentrafo}, 
the subgraph-renormalized
self-energies are related to unrenormalized self-energies according to
\begin{align}
\label{eq:SigmaSR}
\Si^{\FV^\prime\FV}_{\sren,\rT}(k^2,\{c_i\})
={}
\sum_{V_1,V_2=A,Z} Z_{\FV_1\FV^{\prime}}^{1/2} Z_{\FV_2\FV}^{1/2}
\, \Si^{\FV_1\FV_2}_{\rT}(k^2,\{c_{i,0}\}).
\end{align}
Parameter renormalization in the self-energies simply means 
to replace the bare parameters $c_{i,0}$ appearing as arguments
on the r.h.s.\ by the renormalized parameters $c_i$ and 
corresponding renormalization constants $\de c_i$ according to
\begin{align}
\label{eq:parrengeneric}
c_{i,0} = c_i + \de c_i.
\end{align}
Although not needed in the following, but to further classify the
contributing diagrams, we split the contributions
to the unrenormalized self-energy $\Si^{\FV^\prime\FV}_{\rT}(k^2,\{c_{i,0}\})$
into the part $\Si^{\FV^\prime\FV}_{\rT}(k^2,\{c_i\})$ with the
$c_{i,0}$ simply renamed into $c_i$ and a remainder
part $\Si^{\FV^\prime\FV}_{\rT,\de c}(k^2,\{c_i\})$ that
absorbs all effects of the renormalization constants $\de c_i$,
\begin{align}
\Si^{\FV^\prime\FV}_{\rT}(k^2,\{c_{i,0}\}) ={}&
\Si^{\FV^\prime\FV}_{\rT}(k^2,\{c_i\}) +
\Si^{\FV^\prime\FV}_{\rT,\de c}(k^2,\{c_i\})
\nn\\
={}&
\Si^{\FV^\prime\FV}_{\rT,\mathrm{1PI}}(k^2,\{c_i\}) +
\Si^{\FV^\prime\FV}_{\rT,\mathrm{tad}}(k^2,\{c_i\}) +
\Si^{\FV^\prime\FV}_{\rT,\de c}(k^2,\{c_i\}).
\end{align}
The last equation states that $\Si^{\FV^\prime\FV}_{\rT}(k^2,\{c_{i,0}\})$
is calculated from the 1PI and tadpole contributions to the self-energy,
$\Si^{\FV^\prime\FV}_{\rT,\mathrm{1PI}}(k^2,\{c_i\})+
\Si^{\FV^\prime\FV}_{\rT,\mathrm{tad}}(k^2,\{c_i\})$,
which are parametrized by renormalized parameters $c_i$,
and $\Si^{\FV^\prime\FV}_{\rT,\de c}(k^2,\{c_i\})$, resulting from
all possible insertions of counterterm vertices containing
the parameter renormalization constants $\de c_i$.
Note that in this procedure the counterterm contribution
$\Si^{\FV^\prime\FV}_{\rT,\de c}$ does not contain any effects from
field renormalization. 
The effects of field renormalization are completely encoded in the
$Z$~factors appearing on the r.h.s.\ of \refeq{eq:SigmaSR}.
These factors entirely result from the transformation of the
external fields $\FV^\prime$, $\FV$, which are the sources of the 
effective action, in accordance with the well-known fact that
renormalization effects of internal fields in Feynman graphs 
completely cancel between counterterm insertions in propagators
and interaction vertices.
To reduce clutter in the notation, in the following we suppress
the arguments $\{c_{i,0}\}$ and $\{c_i\}$ in the self-energy functions
with the implicit understanding that
$\Si^{\FV^\prime\FV}(k^2) \equiv \Si^{\FV^\prime\FV}(k^2,\{c_{i,0}\})$,
$\Si^{\FV^\prime\FV}_\sren(k^2) \equiv \Si^{\FV^\prime\FV}_\sren(k^2,\{c_i\})$, and
$\Si^{\FV^\prime\FV}_\ren(k^2) \equiv \Si^{\FV^\prime\FV}_\ren(k^2,\{c_i\})$.

Inserting the renormalized vertex functions into the renormalization
conditions $\refeq{eq:GaVArencond}$ and \refeq{eq:dGaVArencond} yields
\begin{align}
\label{eq:GaZArencond}
0 ={}& \Gamma^{\FZ\FA}_{\ren,\rT} (0) =
-\Si^{\FZ\FA}_{\ren,\rT}(0) = 
Z^{1/2}_{\FZ\FZ} Z^{1/2}_{\FZ\FA}(\MZ^2+\de\MZ^2)
-\Si^{\FZ\FA}_{\sren,\rT}(0),
\\
\label{eq:GaAArencond}
0 ={}& 1+\Gamma^{\FA\FA\,\prime}_{\rR,\rT} (0) =
-\Si^{\FA\FA\,\prime}_{\ren,\rT}(0) =
1 - Z_{\FA\FA} - Z_{\FZ\FA} -\Si^{\FA\FA\,\prime}_{\sren,\rT}(0),
\end{align}
where prime means the derivative with respect to the function argument,
i.e.\ $f'(k^2) = \partial f(k^2)/\partial k^2$.
The first of those equations is suited for an order-by-order
calculation of $Z^{1/2}_{\FZ\FA}$, the second provides
$Z_{\FA\FA}$,
\begin{align}
\label{eq:Z_ZA}
Z^{1/2}_{\FZ\FA}
={}& 
\frac{\Si^{\FZ\FA}_{\sren,\rT}(0)}{Z^{1/2}_{\FZ\FZ}\,(\MZ^2+\de\MZ^2)},
\\
\label{eq:Z_AA}
Z_{\FA\FA} ={}&
1 - Z_{\FZ\FA} -\Si^{\FA\FA\,\prime}_{\sren,\rT}(0).
\end{align}
Recalling the leading behaviour of the renormalization constants
in terms of the electromagnetic coupling $\alpha=e^2/(4\pi)$,
\begin{align}
Z_{\FA\FA} = 1+ {\cal O}(\alpha), \quad
Z_{\FZ\FZ} = 1+ {\cal O}(\alpha), \quad
Z^{1/2}_{\FZ\FA} = {\cal O}(\alpha), \quad
Z^{1/2}_{\FA\FZ} = {\cal O}(\alpha), \quad 
\de\MZ^2 = {\cal O}(\alpha),
\end{align}
we see that Eqs.~\refeq{eq:Z_ZA} and \refeq{eq:Z_AA} can be used to 
calculate the $n$-loop contributions to
$Z^{1/2}_{\FZ\FA}$ and $Z_{\FA\FA}$ from the
evaluation of the subgraph-renormalized $ZA$ and $AA$ self-energies
to $n$~loops and from the $(n-1)$-loop contributions to
the renormalization constants $Z_{\FZ\FZ}$ and $\de\MZ^2$ from the Z-boson
sector. Note also that the $(n-1)$-loop contributions to all
parameter renormalization constants $\de c_i$ in general enter the
evaluation of $\Si^{\FV\FA}_{\rT,\de c}$ to $n$~loops.

The above OS renormalization procedure for photons works for any
condition employed to fix the renormalization constants
$Z_{\FZ\FZ}$, $Z_{\FA\FZ}^{1/2}$, and $\de\MZ^2$ of the Z-boson
sector, which enter one loop level lower than intended for 
$Z^{1/2}_{\FZ\FA}$ and $Z_{\FA\FA}$.
We leave the renormalization in the Z-boson sector open, which 
bears additional issues owing to the instability of Z~bosons
(see, e.g., \citere{Denner:2019vbn} and references therein).


Finally, we come back to the stability of the masslessness of the photon
with respect to radiative corrections. To this end, we determine the
location of the particle pole in the photon propagator $G^{AA}_{\mu\nu}$.
Recall that the propagators $G^{b'b}$ for the neutral boson fields $b$, $b'$
(comprising the neutral gauge bosons $A$, $Z$, the neutral Goldstone boson $\chi$,
and the Higgs field $H$ in the SM) result from the
matrix inverse of all two-point vertex functions $-\ri\Gamma^{b'b}$.
The transverse parts $G^{V'V}_{\rT}$ of the neutral-gauge-boson propagators
$G^{V'V}_{\mu\nu}$, which are relevant for the renormalization of the gauge-boson
masses and fields, can be obtained from the matrix inverse of the 
vertex functions $-\ri\Gamma^{V'V}_\rT$ only.
For the $AZ$ system this inversion is simple and leads to the following result
for the transversal part of the unrenormalized photon propagator,
\begin{align}
G^{AA}_\rT(k^2) ={}&  
-\ri \left[k^2+ \Si^{\FA\FA}_{\rT}(k^2) 
- \frac{\left[\Si^{\FA\FZ}_{\rT}(k^2)\right]^2}{k^2-M_{\PZ,0}^2+\Si^{\FZ\FZ}_{\rT}(k^2)}
\right]^{-1}.
\end{align}
The photon, thus, stays massless after switching on the interactions of the
theory if the unrenormalized self-energies obey the relation
\begin{align}
\label{eq:photonmass}
0 ={}&  \Si^{\FA\FA}_{\rT}(0) 
\left[M_{\PZ,0}^2 - \Si^{\FZ\FZ}_{\rT}(0) \right]
+\left[ \Si^{\FA\FZ}_{\rT}(0) \right]^2.
\end{align}
We will check this relation in arbitrary $R_\xi$-gauge and in the BFM below.

Since the propagators are vacuum expectation values of time-ordered products of field
operators, i.e.\ 
$G^{V'V}_{\mu\nu}(x,y) = \langle0| \, T\, V'_{0,\mu}(x)\,V_{0,\nu}(y)\,|0\rangle$,
the transversal parts $G^{V'V}_{\ren,\rT}$ of the renormalized propagators are related to their
unrenormalized counterparts according to
\begin{align}
G^{\FV^{\prime}\FV}_{\rT}(k^2) ={}
\sum_{V_1,V_2=A,Z} Z_{\FV^{\prime}\FV_1}^{1/2} Z_{\FV\FV_2}^{1/2}
G^{\FV_1\FV_2}_{\ren,\rT}(k^2).
\end{align}
Owing to this linear, invertible relation between renormalized and unrenormalized 
propagators, no new poles appear in the set of all $G^{\FV_1\FV_2}_{\ren,\rT}$.
In particular, this and identity~\refeq{eq:photonmass} imply that that the location 
of the pole of $G^{\FA\FA}_{\ren,\rT}$ is at $k^2=0$,
like its unrenormalized counterpart $G^{\FA\FA}_{\rT}$.
That $G^{\FA\FZ}_{\ren,\rT}$ does not develop a pole at $k^2=0$ is achieved by
the renormalization condition~\refeq{eq:GaVArencond} with solution~\refeq{eq:Z_ZA}.
Finally, condition \refeq{eq:GaAArencond} ensures that the residue of
$G^{\FA\FA}_{\ren,\rT}$ for the pole at $k^2=0$ is equal to one.

\subsection{\boldmath{Implications from gauge invariance in arbitrary $R_\xi$-gauge}}

In $R_\xi$-gauge, Green functions, defined by vacuum expectation values of
time-ordered products of field operators, obey ST identities
as a consequence of the Becchi--Rouet--Stora (BRS) symmetry of the Lagrangian after
quantization (see, e.g., 
\citeres{Peskin:1995ev,Weinberg:1996kr,Bohm:2001yx,Schwartz:2013pla}).
These ST identities can be transferred to identities of vertex functions,
known as Lee identities. In general all those identities are very complicated owing to
the occurrence of Green or vertex functions involving Faddeev--Popov fields or
BRS variations of fields.
Fortunately, in order to prove \refeq{eq:photonmass}, we just need a consequence of
the Lee identities for gauge-boson two-point functions of the two-dimensional $AZ$ system,
which can be stated as~\cite{Aoki:1982ed,Bohm:2001yx}
\begin{align}
\label{eq:VVLeeid}
\det\left( \tilde\Gamma^{V'V}_\rL(k^2) \right) = 0,
\end{align}
where $\tilde\Gamma^{V'V}_\rL$ is the longitudinal part of the $V'V$
two-point function from which the tree-level parts of the gauge-fixing terms are subtracted.
Since the full two-point function $\tilde\Gamma^{V'V}_{\mu\nu}$, which is decomposed
as in \refeq{eq:Lorentz_VV}, cannot develop a pole for $k^2\to0$, the $1/k^2$ terms
in the decomposition \refeq{eq:Lorentz_VV} have to cancel for $k^2\to0$, i.e.\
$\tilde\Gamma^{V'V}_\rT(0)=\tilde\Gamma^{V'V}_\rL(0)$ for all $V',V=A,Z$.
Realizing further that $\Gamma^{V'V}_\rT(k^2)=\tilde\Gamma^{V'V}_\rT(k^2)$,
because the tree-level gauge-fixing terms do not contribute to the transversal parts,
Eq.~\refeq{eq:VVLeeid} implies
\begin{align}
\label{eq:VVLeeid2}
\det\left( \Gamma^{V'V}_\rT(0) \right) = 0.
\end{align}
Inserting the decomposition \refeq{eq:GaVVunren} of $\Gamma^{V'V}_\rT$ into lowest-order
parts and self-energies, directly leads to the identity
\refeq{eq:photonmass}, which was to show.

\subsection{Implications from gauge invariance and renormalization constants in the background-field method}

In the BFM, any field $\Psi$ is split into a background part $\hat\Psi$
and a quantum part $\Psi$,
where the quantum fields are the integration variables in the functional integral
used for quantization and the background fields act as sources in the resulting
effective action $\GaBFM[\hat\Psi]$.
The great benefit of the BFM is the invariance of the
effective action $\GaBFM[\hat\Psi]$ under {\it background gauge transformations} of its sources
$\hat\Psi$. 
This leads to QED-like Ward identities for the vertex functions that are
derived from $\GaBFM[\hat\Psi]$ upon taking functional derivatives with respect to
the fields $\hat\Psi$.
These Ward identities imply relations between renormalization constants
similar to the relations known from QED, including the charge renormalization 
constant~\cite{Denner:1994xt}.

In the following, we spell out the procedure of charge renormalization in the BFM,
as suggested in \citere{Denner:1994xt}, based on the
Thomson limit for the $A\bar ff$ vertex for a charged fermion~$f$
with a photon of momentum $k\to0$.
As a result, the charge renormalization constant can be derived from the 
photon wave function renormalization constant, i.e.\ 
the specific fermion~$f$ is not distinguished over any other
charged fermion of the theory. 
The fact that the Thomson limit of the photon coupling to any charged fermion
(and actually to any charged particle) 
can be taken to define the electric unit charge~$e$
in a fully equivalent way proves charge universality.
In the next section we will exploit charge universality, however,
in a different way.

We begin our derivation by recalling the Ward identities for the relevant
{\it unrenormalized} two-point vertex functions for the photon--Z-boson 
system~\cite{Denner:1994xt},
\begin{align}
\label{eq:WIAV}
k^\mu \GaBFM^{\FAhat\FVhat}_{\mu\nu}(k,-k) ={} 0, \qquad V = A,Z.
\end{align}
All definitions and relations for the two-point functions 
$\GaBFM^{\FVhat\FVhat'}_{\mu\nu}$, etc.,
given in \refse{se:VVdefs} carry over to $\GaBFM^{\FVhat\FVhat'}_{\mu\nu}$,
etc., used in that section just by putting hats over $\Ga$ and the fields
$V,V'=A,Z$.
Inserting the decomposition \refeq{eq:Lorentz_VV}
of the two-point functions into Lorentz covariants
into the identities \refeq{eq:WIAV}, we see that the longitudinal parts 
$\GaBFM^{\FAhat\FVhat}_{\rL}$ identically vanish for any $k^2$,
\begin{align}
\GaBFM^{\FAhat\FVhat}_{\rL}(k^{2})=0, \qquad V = A,Z.
\end{align}
In this context, it should be mentioned
that no gauge-fixing terms for the background fields are included yet in the
Lagrangian; those terms provide lowest-order contributions (without any corrections)
to $\GaBFM^{\FAhat\FAhat}_{\rL}$ but not to $\GaBFM^{\FAhat\FAhat}_{\rT}$ 
and are, thus, not entering the
renormalization described in \refse{se:VVdefs}.
Taking into account that two-point functions 
$\GaBFM^{\FVhat^{\prime}\FVhat}_{\mu\nu}$
cannot develop poles in $k^2$, this implies that the transversal parts 
$\GaBFM^{\FAhat\FVhat}_{\rT}$, have to vanish for $k^2=0$,
\begin{align}
\label{eq:GaAVT0}
\GaBFM^{\FAhat\FVhat}_{\rT}(0)=0, \qquad V = A,Z.
\end{align}
For the unrenormalized self-energies at $k^2=0$,
$\Si^{\FAhat\FVhat}_{\rT}(0)$, defined in analogy to \refeq{eq:GaVVunren},
the identity \refeq{eq:GaAVT0}, thus, implies
\begin{align}
\label{eq:SiAVT0}
\Si^{\FAhat\FVhat}_{\rT}(0)=0, \qquad V = A,Z.
\end{align}
Using the analog of \refeq{eq:SigmaSR}, this relation carries over to the subgraph-renormalized
$\FAhat\FZhat$ self-energy,
\begin{align}
\label{eq:BFMSiAZT0SR}
\Si^{\FAhat\FZhat}_{\sren,\rT}(0)
={}&
Z_{\FZhat\FAhat}^{1/2} Z_{\FZhat\FZhat}^{1/2} \, \Si^{\FZhat\FZhat}_{\rT}(0).
\end{align}

Equipped with this identity, the renormalization condition for the 
$\FAhat\FZhat$ vertex function for on-shell photons,
given in \refeq{eq:GaZArencond}, reads
\begin{align}
\label{eq:BFMGaZArencond}
0 ={}& \Gamma^{\FZhat\FAhat}_{\rT} (0) =
-\Si^{\FZhat\FAhat}_{\ren,\rT}(0) = 
Z^{1/2}_{\FZhat\FZhat} Z^{1/2}_{\FZhat\FAhat} \left[ \MZ^2+\de\MZ^2
- \Si^{\FZhat\FZhat}_{\rT}(0) \right].
\end{align}
Since $Z_{\FZhat\FZhat}=1+{\cal O}(\alpha)\ne0$ 
Eq.~\refeq{eq:BFMGaZArencond} implies
\begin{align}
\label{eq:Z_ZAbfm}
Z_{\FZhat\FAhat} = 0.
\end{align}
Using this identity, the result \refeq{eq:Z_AA} for $Z_{\FAhat\FAhat}$ simplifies to
\begin{align}
\label{eq:Z_AAbfm}
Z_{\FAhat\FAhat} ={}&
1 -\Si^{\FAhat\FAhat\,\prime}_{\sren,\rT}(0).
\end{align}

Finally, we observe that condition~\refeq{eq:photonmass}, which implies that the pole
at $k^2=0$ in the photon propagator is not shifted by interactions, is trivially
fulfilled owing to \refeq{eq:GaAVT0} in the BFM.

\section{Charge renormalization and charge universality in the background-field method}
\label{se:chargerenBFM}

The electric unit charged $e$ is renormalized in such a way that the 
fermion--photon interaction for physical (on-shell) fermions does not receive any
correction in the Thomson limit, in which the photon momentum vanishes.
Denoting the relative charge and mass of the fermion~$f$ by $Q_f$ and $m_f$, respectively,
this condition reads
\begin{align}
\label{eq:Affrencond}
\left.\bar{u}(p) \, \GaBFM^{\FA\bar ff}_{\ren,\mu}(0,-p,p) \,
u(p)\right\vert_{p^2=m_f^2}
=-Q_f e \,\bar{u}(p)\gamma_\mu u(p),
\end{align}
where $\GaBFM^{\FA\bar ff}_{\ren,\mu}$ is the renormalized $\FAhat\bar ff$ vertex function
in the BFM and $e$ the renormalized unit charge.
Here $\bar u(p)$ and $u(p)$ are Dirac spinors of the fermion~$f$ with momentum~$p$
fulfilling $p^2=m_f^2$ with the renormalized on-shell mass $m_f$.

In the following, we restore a generation index~$i$ for the considered fermion 
to allow for the possibility of generation mixing.
Since only fermions of the same electric charge can mix, we can keep the
notation $Q_f$ for the common relative charge of the set $\{f_i\}$ of mixing fermions.
The BFM Ward identity for the unrenormalized $\FAhat\bar f_if_j$ 
vertex function reads~\cite{Denner:1994xt} 
(trivially restoring generation indices)
\begin{align}
\label{eq:WIAff}
k^\mu \GaBFM^{\FAhat\Ffbar_i\Ff_j}_{\mu}(k,\bar p, p) ={} -e_0\Qf 
\left[\GaBFM^{\Ffbar_i\Ff_j}(\bar p,-\bar{p}) - \GaBFM^{\Ffbar_i\Ff_j}(-p,p)\right],
\end{align}
where $e_0$ is the bare unit charge and $\GaBFM^{\Ffbar_i\Ff_j}$ 
are the unrenormalized two-point vertex functions of the fermions.
To exploit this identity in the charge renormalization condition~\refeq{eq:Affrencond},
we have to formulate the relations between renormalized and unrenormalized quantities.
To account for the chiral character of the fermions, there are independent sets of
fermionic field renormalization constants $Z^{f,\si}_{ij}$ with $\si=\rR,\rL$
indicating chirality and $i,j$ being matrix indices.
The bare and renormalized fermion fields
$f_{0,i}^\si$ and $f_j^\si$, respectively, are related by
\begin{align}
f_{0,i}^\si = \sum_j \left(Z^{f,\si}_{ij}\right)^{1/2} \, f_j^\si, \qquad
\bar f_{0,i}^\si = \sum_j \left(Z^{f,\si\,*}_{ij}\right)^{1/2} \, \bar f_j^\si, 
\qquad \si=\rR,\rL.
\label{eq:fren}
\end{align}
Together with the field renormalization transformation \refeq{eq:VVfieldren}
of the $\FAhat$ and $\FZhat$ fields, this implies
\begin{align}
\label{eq:ffrentrafo}
\GaBFM^{\Ffbar_i\Ff_j}_{\ren,\mu}(-p,p) ={}&
\sum_{l,n}
\left({Z^{f,\si}_{li}}^*\right)^{1/2} \,
\left(Z^{f,\si}_{nj}\right)^{1/2} \,
\GaBFM^{\Ffbar_l\Ff_n}_{\mu}(-p,p),
\\
\label{eq:Affrentrafo}
\GaBFM^{\FAhat\Ffbar_i\Ff_j}_{\ren,\mu}(k,\bar p, p) ={}&
\sum_{\FVhat=\FAhat,\FZhat} \sum_{l,n}
Z_{\FVhat\FAhat}^{1/2} \, \left({Z^{f,\si}_{li}}^*\right)^{1/2} \,
\left(Z^{f,\si}_{nj}\right)^{1/2} \,
\GaBFM^{\FVhat\Ffbar_l\Ff_n}_{\mu}(k,\bar p, p).
\end{align}
Making use of $Z_{\FZhat\FAhat}=0$ from \refeq{eq:Z_ZAbfm} and introducing the
charge renormalization constant $Z_e$ as ratio between bare charge~$e_0$
and renormalized charge~$e$,
\begin{align}
\label{eq:Ze}
e_0=Z_e e 
\end{align}
delivers the analog of Ward identity \refeq{eq:WIAff} for renormalized 
quantities,
\begin{align}
\label{eq:WIAffren}
k^\mu \GaBFM^{\FAhat\Ffbar_i\Ff_j}_{\ren,\mu}(k,\bar p, p) ={}
-e\Qf \, Z_e \,
Z_{\FAhat\FAhat}^{1/2} \,
\left[\GaBFM^{\Ffbar_i\Ff_j}_\ren(\bar p,-\bar{p}) 
- \GaBFM^{\Ffbar_i\Ff_j}_\ren(-p,p)\right].
\end{align}
This identity is valid for arbitrary momenta $k$, $\bar p$, $p$ obeying momentum
conservation $k+\bar p+p=0$.
Expanding it for $k\to0$ and keeping $p$ fixed, the terms linear in
$k$ obey the relation
\begin{align}
\label{eq:GaBFMAffren}
\GaBFM^{\FAhat\Ffbar_i\Ff_j}_{\ren,\mu}(0,-p, p) ={}&
-e\Qf \, Z_e \,
Z_{\FAhat\FAhat}^{1/2} \,
\frac{\partial\GaBFM^{\Ffbar_i\Ff_j}_\ren(-p,p)}{\partial p^\mu}. 
\end{align}

At this point, we are almost done; we just have to apply Dirac spinors
to $\GaBFM^{\FAhat\Ffbar_i\Ff_j}_{\ren,\mu}(0,-p, p)$ 
from the left and right in \refeq{eq:GaBFMAffren} and to simplify
the term containing $\partial\GaBFM^{\Ffbar_i\Ff_j}_\ren/\partial p^\mu$ 
on the r.h.s.. Note also that we only need the case 
$f=f_i=f_j$ in this last step.
Decomposing the renormalized fermionic two-point function into
Lorentz covariants according to
\begin{align}
\GaBFM^{\Ffbar\Ff}_\ren(-p,p) ={}
    \sum_\si \dsl{p} \omega_\si \GaBFM_\ren^{\Ffbar\Ff,\rV,\si}(p^{2})
    + \sum_\si \omega_\si \GaBFM_\ren^{\Ffbar\Ff,\rS,\si}(p^{2})
\end{align}
with the chirality projectors $\omega_\pm=(1\pm\ga_5)/2$,
it is straightforward to evaluate
$\bar u(p)\,[\partial\GaBFM^{\Ffbar\Ff}_\ren/\partial p^\mu] \,u(p)$
for an on-shell fermion~$f$.
Using some Dirac algebra (Dirac equation, Gordon identity,
$\bar u(p)\ga_5 u(p)=0$), we obtain 
\begin{align}
\bar u(p)\,\frac{\partial\GaBFM^{\Ffbar\Ff}_\ren(-p,p)}{\partial p^\mu}\,u(p) ={}&
    \sum_\si \bar u(p)\,\ga_\mu \omega_\si \,u(p) \,
	\GaBFM_\ren^{\Ffbar\Ff,\rV,\si}(m_f^2)
\\ &{}
    + \sum_\si 2p_\mu\, \bar u(p)\,\omega_\si \,u(p) \, 
	\left[ m_f \GaBFM_\ren^{\Ffbar\Ff,\rV,\si\,\prime}(m_f^2)
	+ \GaBFM_\ren^{\Ffbar\Ff,\rS,\si\,\prime}(m_f^2) \right]
\nn\\ 
={} &
\bar u(p)\,\ga_\mu \,u(p) \,    \sum_\si \left[
	\textstyle\frac{1}{2} \GaBFM_\ren^{\Ffbar\Ff,\rV,\si}(m_f^2)
	+ m_f^2 \GaBFM_\ren^{\Ffbar\Ff,\rV,\si\,\prime}(m_f^2)
	+ m_f \GaBFM_\ren^{\Ffbar\Ff,\rS,\si\,\prime}(m_f^2) \right].
\nn
\end{align}
The factor $\sum_\si [\dots]$ in the last line is easily recognized as the usual
fermionic wave function renormalization factor which is
renormalized to unity by the OS renormalization condition 
(see, e.g., \citere{Denner:2019vbn})
\begin{align}
\lim_{p^2 \to m_f^2} \frac{\dsl{p}+m_f}{p^2-m_f^2}
\left[\GaBFM^{\Ffbar\Ff}_{\ren} (-p,p)\right] u(p) ={}& u(p) 
\end{align}
for the fermion field~$f$. Thus, we finally have
\begin{align}
\label{eq:uGaBFMu}
\bar u(p)\,\frac{\partial\GaBFM^{\Ffbar\Ff}_\ren(-p,p)}{\partial p^\mu}\,u(p) ={}&
\bar u(p)\,\ga_\mu \,u(p).
\end{align}
In summary, combining the charge renormalization condition
\refeq{eq:Affrencond} with \refeq{eq:GaBFMAffren} and \refeq{eq:uGaBFMu}
leads to the simple equation~\cite{Denner:1994xt}
\begin{align}
\label{eq:ZeBFM}
Z_e = Z_{\FAhat\FAhat}^{-1/2}
\end{align}
in the BFM, which is formally identical to the well-known relation 
in QED.
The fact that all dependences from the fermion~$f$, which was used to formulate
the charge renormalization condition in the Thomson limit,
has disappeared in this result for $Z_e$ proves charge universality.
Moreover, Eq.~\refeq{eq:ZeBFM} shows that 
\begin{align}
e_0 \FAhat_{0,\mu}(x) = e \FAhat_\mu(x),
\end{align}
i.e.\ that the product of electromagnetic coupling and background
photon field is not renormalized, again in analogy to a QED relation.

\section{\boldmath{Charge renormalization in arbitrary $R_\xi$-gauge}}
\label{se:chargerenrxi}

\begin{figure}
\centerline{
\raisebox{5em}{(a)}
\includegraphics[scale=1]{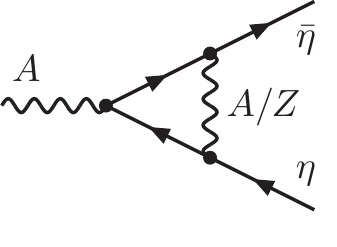} \qquad
\raisebox{5em}{(b)}
\includegraphics[scale=1]{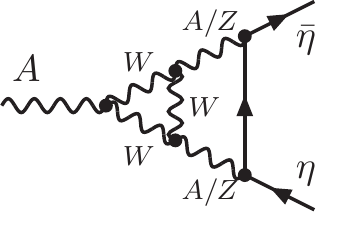} \qquad
\raisebox{5em}{(c)}
\includegraphics[scale=1]{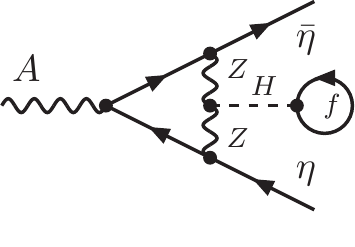}
}
\vspace*{.5em}
\centerline{
\raisebox{4em}{(d)}
\includegraphics[scale=1]{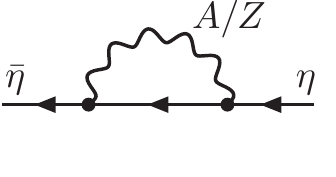} \qquad
\raisebox{4em}{(e)}
\includegraphics[scale=1]{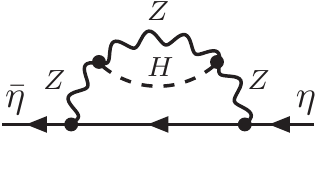} \qquad
\raisebox{4em}{(f)}
\includegraphics[scale=1]{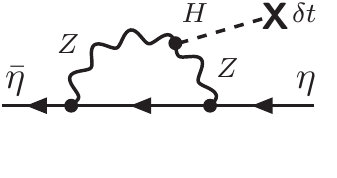}
}
\vspace*{-1.5em}
\caption{Some higher-order diagrams contributing to the unrenormalized 
vertex functions
$\Gamma^{\FV\bar\eta\eta}_{\mu}$ (graphs a--c) and 
$\Gamma^{\bar\eta\eta}$ (graphs d--f), which receive
contributions from 1PI diagrams (graphs a, b, d,~e),
from explicit tadpole diagrams (graph c), and from diagrams
involving tadpole counterterms $\de t$ (graph f).}
\label{fig:Aetaeta-etaeta-diags}
\end{figure}

We extend the considered model, which is the SM or any gauge theory with the gauge group
SU(2)$_\rw\times$U(1)$_Y$ and the same
symmetry-breaking pattern as the SM in the electroweak sector,
by a fermion field~$\eta$ with vanishing weak isopsin, $I_{\rw,\eta}^a=0$, 
and weak hypercharge $Y_{\rw,\eta}$, i.e.\ with electric
charge $Q_\eta=Y_{\rw,\eta}/2$, which is taken as free parameter.
Taking eventually the limit $Q_\eta\to0$, the fermion $\eta$ decouples
from all other particles, and we recover the original theory.
After introducing the field $\eta$, the Lagrangian ${\cal L}$ of the
model is modified to ${\cal L}+{\cal L}_\eta$ with 
\begin{align}
\label{eq:Leta}
{\cal L}_\eta = 
\bar{\eta} \Bigl( \ri \dsl{\partial} 
- {\textstyle\frac{1}{2}} g_1 Y_{\rw,\eta}\dsl{B} -m_\eta\Bigr) \eta =
\overline{\eta} \left[ \ri \dsl{\partial} 
- Q_\eta e \left(\dsl{A}+\frac{\sw}{\cw}\dsl{Z}\right)
 -m_\eta\right] \eta,
\end{align}
with $m_\eta$ denoting the mass of the fermion $\eta$.
Since the field $\eta$ is assumed to be non-chiral, its mass term
is gauge invariant and need not be introduced via the Higgs mechanism.
The non-chirality of $\eta$ also protects us from introducing
anomalies in the model extension.
As in the SM, $g_1$ is the U(1)$_Y$ gauge coupling, $B^\mu$ the U(1)$_Y$ gauge field,
and $\sw=\sin\theta_\rw$ and $\cw=\cos\theta_\rw$ the sine and cosine 
of the weak mixing angle~$\theta_\rw$.
The extra phase symmetry of ${\cal L}_\eta$ with respect to
$\eta\to\re^{\ri\zeta}\eta$ ($\zeta$ real, but arbitrary)
implies that the fermion $\eta$ is stable.
Note that for generic values of $Q_\eta$ the Lagrangian
${\cal L}_\eta$ is the only renormalizable SM extension
of ${\cal L}$ containing the field $\eta$, but no other new field.%
\footnote{If a more general SU(2)$_\rw\times$U(1)$_Y$ theory is considered that contains also
singlet scalars $S_i$, the scalars $S_i$ may also couple to $\eta$ via Yukawa couplings 
$y_i S_i \bar\eta\eta$. The free parameters $y_i$ can be taken to be
infinitesimally small in analogy to $Q_\eta\to0$, so that decoupling of $\eta$
is guaranteed and the arguments below remain valid with obvious minor modifications,
as also detailed in \refse{se:BSM}.}
The renormalization of the model starts by considering all parameters
and fields in the Lagrangian ${\cal L}+{\cal L}_\eta$, with ${\cal L}_\eta$ as given in
\refeq{eq:Leta}, as bare (i.e.\ by adding suffixes ``0'' everywhere).
Owing to charge universality, as proven in the previous section,
we can now take the Thomson limit of the $\FA\bar\eta\eta$ vertex
to define the renormalized electric unit charge~$e$. To this end, we
demand
\begin{align}
\label{eq:Aetaetarencond}
\left.\bar{u}(p) \, \Gamma^{\FA\bar\eta\eta}_{\ren,\mu}(0,-p,p) \,
u(p)\right\vert_{p^2=m_\eta^2}
=-Q_\eta e \,\bar{u}(p)\gamma_\mu u(p)
\end{align}
for the renormalized $\FA\bar\eta\eta$ vertex function
$\Gamma^{\FA\bar\eta\eta}_{\ren,\mu}$ sandwiched between 
Dirac spinors $\bar u(p)$, $u(p)$ of fermions $\eta$ with momentum~$p$
and zero-momentum transfer of the photon.
Here, $m_\eta$ is the renormalized on-shell mass of $\eta$.
The relation between $\Gamma^{\FA\bar\eta\eta}_{\ren,\mu}$
and its bare counterpart $\Gamma^{\FA\bar\eta\eta}_{\mu}$ 
follows from the field renormalization transformation 
for $\eta$,
\begin{align}\label{eq:etafieldren}
\eta_0 = Z^{1/2}_\eta \, \eta, 
\end{align}
and \refeq{eq:VVfieldren} for the photon--Z-boson system
and reads
\begin{align}
\label{eq:GaAetaetaren}
\Gamma^{\FA\bar\eta\eta}_{\ren,\mu}(k,\bar p,p) = 
Z_\eta Z_{\FA\FA}^{1/2} \, \Gamma^{\FA\bar\eta\eta}_{\mu}(k,\bar p,p)  
+ Z_\eta Z_{\FZ\FA}^{1/2} \, \Gamma^{\FZ\bar\eta\eta}_{\mu}(k,\bar p,p).
\end{align}
The bare vertex functions $\Gamma^{\FV\bar\eta\eta}_{\mu}$ 
($\FV=\FA,\FZ$) receive lowest-order contributions
and bare vertex corrections $\Lambda^{\FV\bar\eta\eta}_{\mu}$, 
which consist of 1PI loop diagrams and tadpole corrections,
\begin{align}
\Gamma^{\FA\bar\eta\eta}_{\mu}(k,\bar p,p) ={}& 
- Q_\eta e_0\gamma_\mu +
e_0 \Lambda^{\FA\bar\eta\eta}_{\mu}(k,\bar p,p),
\\
\Gamma^{\FZ\bar\eta\eta}_{\mu}(k,\bar p,p) ={}& 
- Q_\eta e_0\frac{s_{\rw,0}}{c_{\rw,0}}\gamma_\mu +
e_0 \Lambda^{\FZ\bar\eta\eta}_{\mu}(k,\bar p,p),
\end{align}
with $s_{\rw,0}$ and $c_{\rw,0}$ denoting the sine and 
cosine of the bare weak mixing angle.
The important observation is now that all diagrammatic contributions
to $\Lambda^{\FV\bar\eta\eta}_{\mu}$ involve at least
two couplings of photons or Z~bosons to the $\eta$~line that passes through the
whole diagram. 
Some sample diagrams are shown in \reffi{fig:Aetaeta-etaeta-diags}.
For 1PI diagrams it is obvious that at least two couplings to
the $\eta$ line exist, for diagrams with tadpole loops or tadpole counterterms
the same holds true, because
the Higgs field $H$ does not couple to $\eta$.
Since both the photon and the Z~boson couple to $\eta$
proportional to $Q_\eta$, this means that 
$\Lambda^{\FV\bar\eta\eta}_{\mu}={\cal O}(Q_\eta^2)$.
Similarly, all diagrams contributing to the $\eta$~field renormalization
constant $Z_\eta$ involve at least
two couplings of photons or Z~bosons to the $\eta$~line, so that
$Z_\eta=1+{\cal O}(Q_\eta^2)$.
Inserting, thus, $\Gamma^{\FA\bar\eta\eta}_{\ren,\mu}$ from
\refeq{eq:GaAetaetaren} into condition \refeq{eq:Aetaetarencond}
and keeping only terms linear in $Q_\eta$ for $Q_\eta\to0$,
we get
\begin{align}
\label{eq:Aetaetarencond2}
-Q_\eta e \,\bar{u}(p)\gamma_\mu u(p) ={}&
\left.\bar{u}(p) \, Z_\eta \left[
Z_{\FA\FA}^{1/2} \, \Gamma^{\FA\bar\eta\eta}_{\mu}(0,-p,p)  
+ Z_{\FZ\FA}^{1/2} \, \Gamma^{\FZ\bar\eta\eta}_{\mu}(0,-p,p)
\right] u(p)\right\vert_{p^2=m_\eta^2}
\nn\\
={}&
- Q_\eta e_0 \left[
Z_{\FA\FA}^{1/2}  + Z_{\FZ\FA}^{1/2} \frac{s_{\rw,0}}{c_{\rw,0}} \right]
\bar{u}(p) \gamma_\mu u(p) \,+\, {\cal O}(Q_\eta^2).
\end{align}
This relation immediately implies
\begin{align}
\label{eq:eren}
e = e_0 \left[
Z_{\FA\FA}^{1/2}  + Z_{\FZ\FA}^{1/2} \frac{s_{\rw,0}}{c_{\rw,0}} \right],
\end{align}
which is the desired relation between $e$ and $e_0$.
Defining the renormalization constant $Z_e$ as in \refeq{eq:Ze}
and $\de\cw^2$ according to
\begin{align}
c_{\rw,0}^2  = 1- s_{\rw,0}^2 = \cw^2+\de\cw^2 = 1-\sw^2+\de\cw^2,
\end{align}
we can determine $Z_e$ from \refeq{eq:eren},
\begin{align}
Z_e = \left[
Z_{\FA\FA}^{1/2}  + Z_{\FZ\FA}^{1/2} 
\sqrt{\frac{\sw^2-\de\cw^2}{\cw^2+\de\cw^2}} \right]^{-1}.
\end{align}
This is fully equivalent to the result quoted and used in
\citeres{Bauberger:1997zz,Freitas:2002ja,Awramik:2002vu}.

\section{Generalization to non-standard gauge groups}
\label{se:BSM}

The concepts of OS renormalization of the photon field,
of charge universality, and of charge renormalization as described in the
previous sections can be generalized easily to electroweak gauge groups of the type
U(1)$_Y{\times}G$, where $G$ is any Lie group of rank~$r$ (not necessarily
simple or semisimple) and the U(1)$_Y$ group factor plays the analogous
role of weak hypercharge in the SM. 
More precisely, we mean by this that the U(1)$_{\mathrm{em}}$ subgroup of electromagnetic 
gauge transformations mixes transformations of U(1)$_Y$ and $G$, so that
the photon field $A^\mu$ is a non-trivial linear combination of the 
U(1)$_Y$ gauge field $B^\mu$ and the gauge fields $C_k^\mu$ ($k=1,\dots,r$)
of $G$ corresponding to the diagonal group generators in the Lie algebra of $G$.
The mechanism of electroweak symmetry breaking in the considered gauge theory
is widely general, we only assume that electromagnetic gauge invariance
is unbroken.
Specific types of such models are, for instance, described in 
\citere{Babu:1997st} and \citere{Pisano:1991ee} with gauge groups
U(1)$\times$SU(2)$\times$U(1) and U(1)$\times$SU(3), respectively.
If $G$ involves explicit U(1) group factors, the kinetic Lagrangian for the
gauge fields in general includes mixing terms 
$\propto B_{\mu\nu}C_k^{\mu\nu}$ with $B^{\mu\nu}$ and $C_k^{\mu\nu}$
representing the corresponding (gauge-invariant) U(1) field-strength tensors.

Since the generalization of the previous sections to the considered class
of models is straightforward, we can restrict our presentation to the 
salient steps. 
We first have to quantify the transformation of the original gauge fields
$B^\mu$ and $\{C_k^\mu\}$
to (canonically normalized) fields that correspond to mass
eigenstates:
\begin{align}
\begin{pmatrix} B^\mu \\ C_1^\mu \\ \vdots \\ C_r^\mu \end{pmatrix}
= R \begin{pmatrix} A^\mu \\ Z_1^\mu \\ \vdots \\ Z_r^\mu \end{pmatrix},
\qquad
R = \begin{pmatrix} R_{BA} & R_{BZ_1} & \cdots &  R_{BZ_r} \\
R_{C_1A} & R_{C_1Z_1} & \cdots &  R_{C_1Z_r} \\
\vdots & \vdots & \ddots & \vdots \\
R_{C_rA} & R_{C_rZ_1} & \cdots &  R_{C_rZ_r} \end{pmatrix}.
\end{align}
Here, the fields $Z_k^\mu$ ($k=1,\dots,r$) describe neutral massive
gauge bosons similar to the Z~boson of the SM, and the matrix $R$ is a
generalization of the SM rotation matrix parametrized by the weak mixing angle.
Note that $R$ is not necessarily orthogonal or unitary, in particular
in the presence of kinetic mixing among the original gauge fields.

\myparagraph{On-shell renormalization in the photon--Z-boson sector}

Marking bare fields and parameters again with suffixes ``0'', we 
parametrize the field renormalization transformation as follows,
\begin{align}
\begin{pmatrix} A_0^\mu \\ Z_{0,1}^\mu \\ \vdots \\ Z_{0,r}^\mu \end{pmatrix}
= \begin{pmatrix} Z^{1/2}_{AA} & Z^{1/2}_{AZ_1} & \cdots &  Z^{1/2}_{AZ_r} \\
Z^{1/2}_{Z_1A} & Z^{1/2}_{Z_1Z_1} & \cdots &  Z^{1/2}_{Z_1Z_r} \\
\vdots & \vdots & \ddots & \vdots \\
Z^{1/2}_{Z_rA} & Z^{1/2}_{Z_rZ_1} & \cdots &  Z^{1/2}_{Z_rZ_r} \end{pmatrix}
\begin{pmatrix} A^\mu \\ Z_1^\mu \\ \vdots \\ Z_r^\mu \end{pmatrix},
\end{align}
similar to \refeq{eq:VVfieldren} in the SM.
Making use of the same definitions for vertex functions and self-energies
as in \refse{se:VVdefs},
the renormalized transversal parts of the two-point functions of neutral gauge bosons
are given by
\begin{align}
\label{eq:GaVVrenBSM}
\Gamma^{\FV^\prime\FV}_{\rR,\rT} (k^2) 
={}&
-\de_{\FV^\prime\FV} \, (k^2 -\de_{\FV\FZ_k} \,M_{\PZ_k}^2)
-\Si^{\FV^\prime\FV}_{\ren,\rT}(k^2)
\nn\\
={}&
- Z^{1/2}_{\FA\FV^\prime} Z^{1/2}_{\FA\FV} k^2
- \sum_k Z^{1/2}_{\FZ_k\FV^\prime} Z^{1/2}_{\FZ_k\FV}(k^2-M_{\PZ_k}^2-\de M_{\PZ_k}^2)
-\Si^{\FV^\prime\FV}_{\sren,\rT}(k^2),
\end{align}
with $V,V'=A,Z_1,\dots,Z_r$.
The OS renormalization conditions \refeq{eq:GaZArencond} and 
\refeq{eq:GaAArencond} for external photons analogously hold for the
$AA$ and all $Z_k A$ vertex functions and imply
\begin{align}
\label{eq:GaZArencondBSM}
0 ={}& \Gamma^{\FZ_k\FA}_{\ren,\rT} (0) =
-\Si^{\FZ_k\FA}_{\ren,\rT}(0) = 
\sum_l Z^{1/2}_{\FZ_l\FZ_k} Z^{1/2}_{\FZ_l\FA}(M_{\PZ_l}^2+\de M_{\PZ_l}^2)
-\Si^{\FZ_k\FA}_{\sren,\rT}(0),
\\
\label{eq:GaAArencondBSM}
0 ={}& 1+\Gamma^{\FA\FA\,\prime}_{\rR,\rT} (0) =
-\Si^{\FA\FA\,\prime}_{\ren,\rT}(0) =
1 - Z_{\FA\FA} - \sum_k Z_{\FZ_k\FA} -\Si^{\FA\FA\,\prime}_{\sren,\rT}(0).
\end{align}
Similar to \refeq{eq:Z_ZA} and \refeq{eq:Z_AA} in the SM, these conditions can be
used to calculate the renormalization constants $Z_{\FZ_k\FA}$ and
$Z_{\FA\FA}$ recursively order by order from the relations
\begin{align}
\label{eq:Z_ZABSM}
Z^{1/2}_{\FZ_k\FA}
={}& 
\frac{\Si^{\FZ_k\FA}_{\sren,\rT}(0)-\sum_{l\,(l\ne k)} 
  Z^{1/2}_{\FZ_l\FZ_k} Z^{1/2}_{\FZ_l\FA}
  (M_{\PZ_l}^2+\de M_{\PZ_l}^2)}{Z^{1/2}_{\FZ_k\FZ_k}\,(M_{\PZ_k}^2+\de M_{\PZ_k}^2)},
\\
\label{eq:Z_AABSM}
Z_{\FA\FA} ={}&
1 - \sum_k Z_{\FZ_k\FA} -\Si^{\FA\FA\,\prime}_{\sren,\rT}(0),
\end{align}
since the
leading behaviour of the occurring renormalization constants is given by
\begin{align}
Z_{\FA\FA} = 1+ {\cal O}(\alpha), \quad
Z^{1/2}_{\FZ_k\FZ_l} = \de_{kl}+ {\cal O}(\alpha), \quad
Z^{1/2}_{\FZ_k\FA} = {\cal O}(\alpha), \quad
Z^{1/2}_{\FA\FZ_k} = {\cal O}(\alpha), \quad 
\de M_{\PZ_k}^2 = {\cal O}(\alpha).
\end{align}
For calculating $Z_{\FZ_k\FA}$ and $Z_{\FA\FA}$ at the $n$-loop level, the 
field renormalization constants $Z^{1/2}_{\FZ_k\FZ_l}$ and the mass
renormalization constants $\de M_{\PZ_k}^2$ for the Z$_k$ boson are only required
to the $(n-1)$-loop level.

\myparagraph{Photon-field renormalization, charge renormalization, and 
charge universality in the BFM}

The invariance of the BFM effective action w.r.t.\ electromagnetic 
(background) gauge transformation implies the validity of the
Eqs.~\refeq{eq:WIAV}--\refeq{eq:SiAVT0} for all $\hat V=\hat A,\hat Z_1,\dots,\hat Z_r$.
Although \refeq{eq:BFMSiAZT0SR} and \refeq{eq:BFMGaZArencond} now involve sums
of all fields $Z_k$, induction in the loop order $n$ can be applied to show
\begin{align}
\label{eq:Z_VAbfmBSM}
Z_{\FZhat_k\FAhat} = 0, \qquad k=1,\dots,r,
\end{align}
and, thus, also
\begin{align}
Z_{\FAhat\FAhat} =
1 -\Si^{\FAhat\FAhat\,\prime}_{\sren,\rT}(0).
\end{align}
With these results, the whole derivation of $Z_e$ described in 
\refse{se:chargerenBFM} for the BFM goes through with the only
modification of extending some sums over $\hat V=\hat A,\hat Z$
to sums over $\hat V=\hat A,\hat Z_1,\dots,\hat Z_r$.
As a result, the charge renormalization constant $Z_e$ is
given by $Z_e = Z_{\FAhat\FAhat}^{-1/2}$ as in \refeq{eq:ZeBFM}.
This again proves charge universality in the model, independent of the
use of the BFM in the proof.

\myparagraph{Charge universality in $R_\xi$-gauge}

To exploit charge universality in the determination of $Z_e$ in arbitrary
$R_\xi$-gauge, we again introduce a fake fermion $\eta$ with the same properties
as described in \refse{se:chargerenrxi}, i.e.\
$\eta$ only carries infinitesimal hypercharge $Y_{\rw,\eta}$, but no 
non-trivial quantum number of $G$.
If the model contains 
singlet scalars $S_i$, the scalars $S_i$ may couple to $\eta$ via Yukawa couplings.
The corresponding couplings $y_i$ are free parameters of the model and
can be taken to be
infinitesimally small in analogy to $Q_\eta\to0$, so that decoupling of $\eta$
is guaranteed.
The Lagrangian $\L_\eta$, thus, reads
\begin{align}
\label{eq:LetaBSM}
{\cal L}_\eta = {} &
\bar{\eta} \Bigl( \ri \dsl{\partial} 
- {\textstyle\frac{1}{2}} g_1 Y_{\rw,\eta}\dsl{B} -m_\eta
-\sum_i y_i S_i \Bigr) \eta 
\nn\\
= {} &\overline{\eta} \bigg[ \ri \dsl{\partial} 
- Q_\eta e
\bigg( \dsl{A}+\sum_k \frac{R_{BZ_k}}{R_{BA}} \dsl{Z}_k \bigg) -m_\eta
-\sum_i y_i S_i \bigg] \eta,
\end{align}
where we have identified 
\begin{align}
\textstyle\frac{1}{2} Y_{\rw,\eta} = Q_\eta, \qquad
g_1 R_{BA} = e.
\end{align}
Following the same reasoning as in \refse{se:chargerenrxi}, the
renormalized $\FA\bar\eta\eta$ vertex function is given by
\begin{align}
\label{eq:GaAetaetarenBSM}
\Gamma^{\FA\bar\eta\eta}_{\ren,\mu}(k,\bar p,p) = 
Z_\eta Z_{\FA\FA}^{1/2} \, \Gamma^{\FA\bar\eta\eta}_{\mu}(k,\bar p,p)  
+ \sum_k Z_\eta Z_{\FZ_k\FA}^{1/2} \, \Gamma^{\FZ_k\bar\eta\eta}_{\mu}(k,\bar p,p),
\end{align}
with the 
unrenormalized $\FV\bar\eta\eta$ vertex functions
\begin{align}
\Gamma^{\FA\bar\eta\eta}_{\mu}(k,\bar p,p) ={}& 
- Q_\eta e_0\gamma_\mu +
e_0 \Lambda^{\FA\bar\eta\eta}_{\mu}(k,\bar p,p),
\\
\Gamma^{\FZ_k\bar\eta\eta}_{\mu}(k,\bar p,p) ={}& 
- Q_\eta e_0\,\frac{R_{0,BZ_k}}{R_{0,BA}}\,\gamma_\mu +
e_0 \Lambda^{\FZ_k\bar\eta\eta}_{\mu}(k,\bar p,p).
\end{align}
Again the vertex corrections $\Lambda^{\FA\bar\eta\eta}_{\mu}$ and
$\Lambda^{\FZ_k\bar\eta\eta}_{\mu}$ as well as the field renormalization constant
$\de Z_\eta = Z_\eta-1$ receive only corrections that are suppressed at least
by quadratic factors in the new couplings, such as $Q_\eta^2$ or $Q_\eta y_i$.
Typical diagrams contributing to those corrections at the order ${\cal O}(Q_\eta^2)$
(or higher in $Q_\eta$)
can be obtained from the graphs shown in \reffi{fig:Aetaeta-etaeta-diags}
upon interpreting the field $Z$ as any of the $Z_k$ and taking the Higgs
field $H$ as a any Higgs field of the model.
Equation~\refeq{eq:Aetaetarencond2} then generalizes to the considered model
in an obvious way, and we obtain the final result for the 
charge renormalization constant:
\begin{align}
\label{eq:ZeBSM}
Z_e = \left[
Z_{\FA\FA}^{1/2}  + \sum_k Z_{\FZ_k\FA}^{1/2} \,
\frac{R_{BZ_k}+\de R_{BZ_k}}{R_{BA}+\de R_{BA}} \right]^{-1}.
\end{align}
Here we have formally introduced renormalization constants $\de R_{BA}$ and
$\de R_{BZ_k}$ for the matrix elements of $R$ according to 
$R_0=R+\de R$, but we have to keep in mind that not all those
constants are independent, because not all elements of $R$ are independent
free parameters of the theory.
Finally, we specialize \refeq{eq:ZeBSM} to the one-loop level,
which is sufficient for most applications.
To this end, we expand the charge and field renormalization constants according to
\begin{align}
Z_e = 1+\de Z_e +{\cal O}(\alpha^2), \qquad
Z_{\FA\FA}^{1/2} = 1+\textstyle\frac{1}{2} \de Z_{\FA\FA}+{\cal O}(\alpha^2), \qquad
Z_{\FZ_k\FA}^{1/2} = \textstyle\frac{1}{2} \de Z_{\FZ_k\FA}+{\cal O}(\alpha^2)
\end{align}
and find
\begin{align}
\de Z_e = -\frac{1}{2} \de Z_{\FA\FA}  - \frac{1}{2}\sum_k 
\frac{R_{BZ_k}}{R_{BA}} \, \de Z_{\FZ_k\FA}.
\end{align}
At one loop, $Z_e$ is, thus, independent of the renormalization conditions
chosen for the mixing matrix $R$ and for the Z-boson masses $M_{\PZ_k}$.

The case of the SM is trivially recovered from the results of this section
upon identifying
$G={}$SU(2)$_\rw$, $r=1$, $Z_1^\mu=Z^\mu$, $R_{BZ_1}=\sw$, and $R_{BA}=\cw$.

\section{Conclusions}
\label{se:concl}

In this article we have derived an all-order form for the 
renormalization constant $Z_e$ of electric charge, as defined in the Thomson limit,
in an arbitrary $R_\xi$-gauge which expresses $Z_e$ in terms of self-energies
of the photon--Z-boson system only.
We confirm the result that has been given in the literature before, but
the derivations of which are either tied to specific gauges, restricted
to the two-loop level, or even contain inconsistencies. 
Our derivation, thus, provides an a posteriori justification
for the few calculations of two-loop electroweak corrections based on the 
assumed form for $Z_e$.

Our derivation exploits charge universality, i.e.\ the fact that
the electric unit charge can be defined from the Thomson 
(low-energy/momentum) limit of the
photonic interaction with any charged fermion. 
Charge universality, for instance, follows from the 
known universal form of the charge renormalization constant within the 
background-field formalism, which we have rederived in this paper as well.
Exploiting charge universality, we formulate the
charge renormalization condition for the photonic interaction of a fake
fermion with infinitesimal weak hypercharge and vanishing weak isospin,
which effectively decouples from all other particles.
Without spelling out the details in this paper,
we have repeated the derivation with a fake boson of spin~0 which produces
the same universal result for $Z_e$ as for the fake fermion. Charge universality, thus,
holds for spin-0 bosons too, as expected.

Moreover, we have discussed the derivation of $Z_e$ both in the conventional
quantization formalism for gauge theories and in the background-field method.
Since the determination of $Z_e$ in the Thomson limit of the 
fermion--photon vertex with on-shell fermions and an on-shell photon is based
on the property of an S-matrix element, the result on $Z_e$ has to be independent
of the chosen gauge or quantization procedure. Comparing the explicit results for $Z_e$
obtained via different gauges or quantization procedures order by order, therefore 
provides useful checks on higher-order calculations.

The presented derivation of charge renormalization
only makes use of the gauge structure of the model, but does not depend
on the matter particle content, the Higgs sector, or other properties.
Thus, the result for $Z_e$ as obtained for the SM literally
holds in all spontaneously broken gauge theories with the
SU(2)$_\rw\times$U(1)$_Y$ gauge group and SM-like gauge symmetry breaking
in the electroweak sector.
Finally, we have determined the charge renormalization constant to all 
perturbative orders in the more general class of spontaneously broken gauge theories
with gauge group U(1)$_Y{\times}G$ with any Lie group $G$, only assuming
that electromagnetic gauge symmetry is unbroken and mixes with
U(1)$_Y$ transformations in a non-trivial way.

\appendix
\section*{Appendix}
\section*{\boldmath{Problems in previous all-order determinations of $Z_e$}}
\setcounter{section}{1}

\begin{figure}
\centerline{
\raisebox{5em}{(a)}
\includegraphics[scale=1]{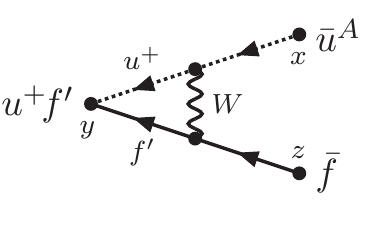} \qquad
\raisebox{5em}{(b)}
\includegraphics[scale=1]{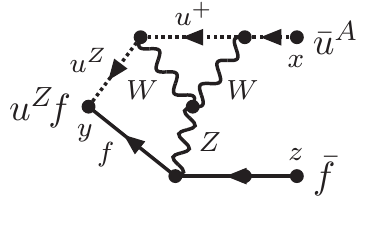} \qquad
\raisebox{5em}{(c)}
\includegraphics[scale=1]{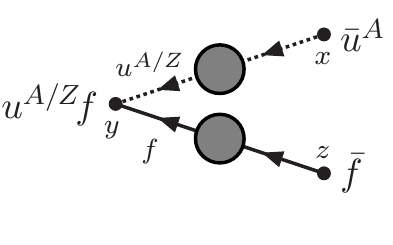}
}
\vspace*{-1.5em}
\caption{Various graphs contributing to
the $\int\rd^D q\,G^{u^V\bar u^A f'\bar f}(q,k,p_1-q,p_2)$ in momentum space,
which correspond to the Green functions 
$G^{u^V\bar u^A f'\bar f}(y,x,y,z)$ in position space:
(a,b) two connected graphs 
and (c) the generic graph for disconnected parts, with the grey blob representing
1PI or tadpole contributions.}
\label{fig:STAff-diags}
\end{figure}

In \citere{Bauberger:1997zz}, the derivation of the charge renormalization constant $Z_e$
starts from the ST identity obtained from the vanishing BRS variation
of the Green function $\langle0|T \,\bar u^A(x)\, \psi_f(y)\,\bar\psi_f(z)\,|0\rangle$.
Translating the corresponding relations in Sects.~4.2 and 4.3 of \citere{Bauberger:1997zz}
to the conventions of \citere{Denner:2019vbn}, this schematically implies
\begin{align}
0 ={} & k^\mu G_{\mu}^{Af\bar f}(k,p_1,p_2)
+ a_Z k^\mu G_{\mu}^{Zf\bar f}(k,p_1,p_2)
+ a_\chi G^{\chi f\bar f}(k,p_1,p_2)
\nn\\
&{}
+ \sum_{V=A,Z} a^Y_{fV} \int\frac{\rd^Dq}{(2\pi)^D} \left[
G^{u^V\bar u^A f'\bar f}(q,k,p_1-q,p_2)
- G^{u^{V^\dagger}\bar u^A f\bar f'}(q,k,p_1,p_2-q)
\right]
\nn\\
&{}
+ \sum_{V=Z,W^\pm} a^\rw_{fV} \int\frac{\rd^Dq}{(2\pi)^D} \left[
\omega_- G^{u^V\bar u^A f'\bar f}(q,k,p_1-q,p_2)
- G^{u^{V^\dagger}\bar u^A f\bar f'}(q,k,p_1,p_2-q) \omega_+
\right],
\label{eq:STAff}
\end{align}
where the constants $a_Z$, $a_\chi$ are determined by the gauge-fixing term
of the photon field. 
The constants $a^Y_{fV}$ and $a^\rw_{fV}$ express the transformation
properties of the fermion~$f$ w.r.t.\ U(1)$_Y$ and SU(2)$_\rw$
gauge transformations, respectively. 
In case $V$ is a charged gauge boson, the field~$f'$
is the field of its weak isospin partner, otherwise $f'=f$.
Figure~\ref{fig:STAff-diags} illustrates some Feynman diagrams and diagram types
contributing to the 
Green functions $G^{u^V\bar u^A f'\bar f}$;
graphs contributing to $G^{u^{V^\dagger}\bar u^A f\bar f'}$ look similar,
with the ghost and fermion lines meeting in the field point at~$z$.
Identity \refeq{eq:STAff}
is correct and certainly bears the desired information on the
Thomson limit of the $Af\bar f$ vertex, however, the reasoning explained in
\citere{Bauberger:1997zz} does not hold:
\begin{itemize}
\item
In order to get rid of the contribution of SU(2)$_\rw$ gauge bosons in the last line of
\refeq{eq:STAff}, the identity is formulated for right-handed fields, and
the mixing between right- and left-handed fields is ignored during the 
amputation of the external propagators and the projection to on-shell states.
This simplification, though, could be dropped by a projection of 
\refeq{eq:STAff} to right-handed chirality from the left and from the right
with a subsequent amputation of the full fermion propagators.
We have carried out this more laborious procedure at one loop.
The oversimplification seems to be no show stopper.
\item
A serious problem, however, concerns the simultaneous on-shell limit
$k\to0$, $p_1^2\to m_f^2$, and $p_2^2\to m_f^2$ after amputation, 
which is presented in \citere{Bauberger:1997zz} in a sketchy way. 
The claim that all connected parts of 
$G^{u^V\bar u^A f'\bar f}$ and 
$G^{u^{V^\dagger}\bar u^A f\bar f'}$ vanish in this limit,
unfortunately does not hold, because this multiple limit
is more subtle.

If we first go on shell with the fermion momenta $p_1$, $p_2$ after amputation, 
leaving $k$ open, in fact the unpleasant connected parts of 
$G^{u^V\bar u^A f'\bar f}$ and 
$G^{u^{V^\dagger}\bar u^A f\bar f'}$ vanish due to a missing pole in one
of the external fermion legs.
The projection of \refeq{eq:STAff} to on-shell fermion states leads to a relation
between the axial-vector and scalar formfactors of the $Af\bar f$ vertex 
and self-energies for arbitrary $k^2$. 
At one loop we have checked this identity by explicit calculation.
For $k^2=0$, this identity was, e.g., given as
Eq.~(3.33) in \citere{Hollik:1988ii},
as Eq.~(3.30) in \citere{Denner:1991kt}, or as
the second relation in Eq.~(C.31) of \citere{Denner:2019vbn}.

However, in order to obtain the required relation for the 
vector formfactor of the $Af\bar f$ vertex, which is, e.g., given
as Eq.~(3.27) in \citere{Hollik:1988ii},
as Eq.~(3.29) in \citere{Denner:1991kt}, or as
the first relation in Eq.~(C.31) of \citere{Denner:2019vbn} at one loop,
the on-shell limit has to be carried out differently:
Only one of the fermion momenta, called $p$ in the following, 
can be set on shell at the beginning,
followed by an derivative w.r.t.\ the photon momentum $k$,
with subsequently taking $k\to0$. With the last step, the second 
fermion line goes on shell automatically.
Note that taking the derivative w.r.t.\ $k$ while keeping $p$ fixed
increases the order of the pole in the propagator of the off-shell fermion,
which has momentum $p+k$. For $k\to0$, there are, thus, poles of order two
at $p^2=m_f^2$ which prevents the connected parts of 
$G^{u^V\bar u^A f'\bar f}$ and
$G^{u^{V^\dagger}\bar u^A f\bar f'}$ from vanishing in the 
on-shell projection of the two external fermion lines in general.

At one loop, however, the mentioned terms still vanish if the fermion
line is projected to right-handed chirality on both fermion legs,
because then no one-loop Feynman diagram exists that links the 
photonic ghost line to the fermion lines.
Beyond one loop, connected graphs contributing 
$G^{u^V\bar u^A f'\bar f}$ and
$G^{u^{V^\dagger}\bar u^A f\bar f'}$ exist (see \reffi{fig:STAff-diags}b), 
and there is no reason
for them to vanish.

\item
Finally, the evaluation of the ghost propagators and their renormalization
in the disconnected contributions to $G^{u^V\bar u^A f'\bar f}$ and
$G^{u^{V^\dagger}\bar u^A f\bar f'}$ in \citere{Bauberger:1997zz}
is not correct. 
Equations~(4.38) and (4.39) of \citere{Bauberger:1997zz} express the required
residues of the $G^{u^A\bar u^A}$ and $G^{u^Z\bar u^A}$ propagators at $k^2=0$
in terms of the field renormalization constants 
$Z_{AA}$ and $Z_{ZA}$, respectively.
However, since $Z_{AA}$ already involves non-vanishing contributions
from closed fermion loops, but $G^{u^A\bar u^A}$ does not, Eq.~(4.38)
of \citere{Bauberger:1997zz} is obviously invalid.

The correct evaluation of the ghost propagators can, e.g., be based on
the BRS invariance of 
the (unrenormalized) Green function $\langle|T \,\bar u^A(x)\, B^\mu(y)\,|0\rangle$.
The resulting ST identity for the ghost propagators expresses
$G^{u^B\bar u^A}$ in terms of the longitudinal part of the
$AZ$~self-energy.
\end{itemize}
With the help of the mentioned corrections,we have successfully derived the
known form of $Z_e$ at the one-loop level, starting from the ST identity
\refeq{eq:STAff}, which provides an alternative to the derivation described in
App.~C of \citere{Denner:2019vbn} based on Lee identities.
We do, however, not see a way how to carry out a corresponding all-order proof
by simple amendments.

Completely independent of the proof based on \refeq{eq:STAff}, it was
suggested in \citere{Bauberger:1997zz} and in App.~A of \citere{Awramik:2002vu}
to deduce $Z_e$ from the fact that the product $g_1 B^\mu$ need not be renormalized.
This fact is justified in those papers upon referring to 
Sect.~3.4.3 of \citere{Aoki:1982ed} where it was deduced from Lee identities
in the course of proving charge universality. Since this derivation in
\citere{Aoki:1982ed} is carried out in the Landau gauge, it is actually not
clear without further justification that no modifications are necessary
in general $R_\xi$-gauge.
At the two-loop level, this missing justification was provided in 
\citeres{Actis:2006ra,Actis:2006rb,Actis:2006rc} where the non-renormalizability
of $g_1 B^\mu$ was first assumed for charge renormalization and
in a second step checked by explicit calculation that all $Af\bar f$ vertex corrections
vanish in the Thomson limit.
A corresponding all-order proof that the non-renormalizability hypothesis of
$g_1 B^\mu$ is equivalent to charge renormalization in the Thomson limit
in arbitrary $R_\xi$-gauge to all orders, to our knowledge, does not exist in the literature.

\section*{Acknowledgements}

Ansgar Denner and Giampiero Passarino are gratefully acknowledged for
discussions on the subject. 
This work is supported via grant DI~785/1 of the Deutsche
Forschungsgemeinschaft (DFG).

\bibliographystyle{jhep}
\providecommand{\href}[2]{#2}\begingroup\raggedright\endgroup

\end{document}